\documentstyle[sprocl,epsfig]{article}

\def\Journal#1#2#3#4{{#1} {\bf #2}, #3 (#4)}

\def\EPJ{{\em Eur. Phys. J.} C}

\def\be{\begin{equation}}
\def\ee{\end{equation}}
\def\bea{\begin{eqnarray}}
\def\eea{\end{eqnarray}}

\def\gevsq{GeV$^2$}

\def\De{\Delta}
\def\csq{\chi^2}
\def\pa{\partial}
\def\pla{p_{\lambda}}
\def\pmu{p_{\mu}}
\def\smu{s_{\mu}}
\def\si{\sigma}
\def\la{\lambda}

\newcommand{\Eq}[1]{Eq.~(\ref{#1})}
\newcommand{\Fi}[1]{Fig.~\ref{#1}}

\newcommand{\ma}[1]{\mbox{\sf #1}}

\begin{document}

\title{ERROR PROPAGATION IN QCD FITS}

\author{M. BOTJE}

\address{NIKHEF, P.O. Box 41882, 
1009DB Amsterdam, The Netherlands\\E-mail: m.botje@nikhef.nl} 


\maketitle\abstracts{The parton momentum density distributions of the
proton were obtained from a NLO QCD analysis of HERA and fixed target
structure function data.  The resulting parton distribution set includes
the full information on errors and correlations.}

\section{Introduction}
  
Standard sets of parton densities~\cite{ref:grv,ref:mrst,ref:cteq} are
widely used to calculate hard scattering cross sections in
hadron-hadron and lepton-hadron collisions. However, none of these
sets give the errors on the parton densities which tend to dominate
the uncertainties on the predicted cross sections.

To make a parton distribution set available which includes the full
information on errors and correlations\footnote{Available from {\sf
http://www.nikhef.nl/user/h24/qcdnum}} we have performed a NLO QCD
analysis of HERA and fixed target structure function
data.~\cite{ref:mbfit} In this contribution we describe how the
experimental errors were propagated in the QCD fit.

\section{Error propagation}

In the fit, the correlated experimental systematic errors were
incorporated in the model prediction of the structure functions:
\be \label{eq:sysdef}
F_i(p,s) = F_i^{\rm QCD}(p) \biggl( 1 + 
           \sum_{\la}  s_{\la} \De_{i \la}^{\rm syst} \biggr)
\ee
where $F_i^{\rm QCD}(p)$ is the QCD prediction and $\De_{i \la}^{\rm syst}$
is the relative systematic error on data point $i$ stemming from source
$\la$. The fitted parameters $p$ describe the parton densities at the
input scale $Q^2_0$ and $s$ denotes the set of systematic parameters.
Assuming that the $s_{\la}$ are uncorrelated and Gaussian distributed with
zero mean and unit variance, the $\csq$ was
defined in the usual way and two Hessian matrices $\ma{M}$ and $\ma{C}$
were evaluated at the minimum $\csq$:
\be \label{eq:hessian}
M_{\la \mu} = \frac{1}{2} \frac{\pa^2 \csq}{\pa \pla \pa \pmu}
,\ \  
C_{\la \mu} = \frac{1}{2} \frac{\pa^2 \csq}{\pa \pla \pa \smu}.
\ee
The statistical and systematic~\cite{ref:pascaud} covariance matrices
were calculated from
\be \label{eq:vstat}
\ma{V}^{\rm stat} = \ma{M}^{-1},\ \ 
\ma{V}^{\rm syst} =  \ma{M}^{-1} \ma{C} \ma{C}^T \ma{M}^{-1}. 
\ee
The error on any function $F$ of the parameters $p$ is, to first order,
given by
\be \label{eq:eprop}
(\De F)^2 = \sum_{\la} \sum_{\mu} \frac{\pa F}{\pa \pla} V_{\la \mu}
\frac{\pa F}{\pa \pmu}
\ee
where $\ma{V}$ is the statistical, the systematic, or if the total error
is to be calculated, the sum of both covariance matrices.

Three additional sources of error were considered in the analysis: (i)
Errors due to the uncertainties on the input parameters ($\alpha_s$
etc.). The error bands are defined as the envelope of the results from
the central fit and two additional fits where each input parameter was
lowered or raised by the error; (ii) An `analysis' error band is
defined as the envelope of the central fit and 10 alternative fits
where, for instance, the cuts on the data were varied; (iii) The
renormalization and factorization scale uncertainties, obtained from
fits where both scales were independently varied in the range $Q^2/2 <
\mu^2 < 2Q^2$.

In \Fi{fig:fig1} 
\begin{figure}[t] 
\center\epsfig{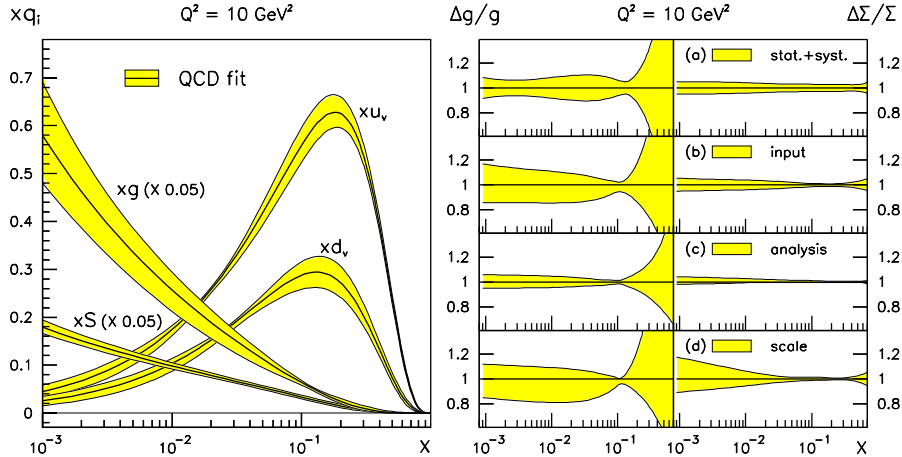}
\caption{(Left) The parton densities from this analysis versus $x$ at $Q^2 =
10$~\gevsq. (Right) The errors on the gluon and quark densities
from the various sources described in the text.
\label{fig:fig1}} 
\end{figure}
we show the parton densities obtained from this analysis (left hand
plot). The error bands correspond to the quadratic sum of all errors
except the scale uncertainties. The relative error contributions to
the gluon density and the singlet quark density are shown in the right
hand plot of \Fi{fig:fig1}. It is seen that the analysis error band is
small. For the gluon density the remaining contributions to the error
are roughly equal in size whereas for the quarks it turns out that the
scale uncertainty is the largest source of error.

\section{Parton distribution set}

Stored on a computer readable file are the statistical and systematic
covariance matrices, the parton densities $f_i$, the derivatives $\pa
f_i / \pa \pla$ (both from the central fit), the results from the
systematic fits (where the input parameters or scales were changed) and
the analysis error band. The kinematic range covered by the parton
densities is $9 \times 10^{-4} < x < 1$ and $1 < Q^2 < 9 \times
10^4$~\gevsq. 

A computer program~\cite{ref:epdflib} gives fast access to these
results and provides tools, which make use of \Eq{eq:eprop}, to
propagate the statistical and systematic errors to any function $F$ of
the parton densities.  As an input to the program the user should
provide a calculation of the derivatives $\pa F / \pa \pla$.  Let us
take as an example a hadron-hadron cross section which can be written
as a convolution of the parton densities and a hard scattering cross
section,
\be \label{eq2:hard}
\si = \sum_{ij} f_i \otimes f_j \otimes \hat{\si}_{ij}.
\ee
To calculate the error on $\si$ it is sufficient to provide a function
which computes the derivatives
\be \label{eq:sighadron}
\frac{\pa \si}{\pa \pla} = \sum_{ij} \left[
\frac{\pa f_i}{\pa \pla} \otimes f_j +
f_i \otimes \frac{\pa f_j}{\pa \pla} \right] \otimes \hat{\si}_{ij}.
\ee
This calculation is straight forward since the $f_i$ and the 
$\pa f_i / \pa \pla$ are available from the input file.

Finally, we remark that the errors from a QCD fit are not determined
by the experimental errors alone but also depend, and maybe quite
strongly, on the assumptions made in the analysis, in particular on
the parameterizations chosen for the parton densities at the input
scale $Q^2_0$.

\end{document}